\documentclass[twocolumn, amsmath, amssymb, aps, prl, showkeys, nolongbibliography, superscriptaddress, numerical]{revtex4-2}
\usepackage{graphicx}% Include figure files
\usepackage{dcolumn}% Align table columns on decimal point
\usepackage{bm}% bold math
\usepackage{hyperref}
\usepackage{lipsum}
\usepackage{booktabs}
\usepackage[mathlines]{lineno}% Enable numbering of text and display math
\usepackage{natbib}

\usepackage{xcolor}

\begin{document}

\preprint{APS/123-QED}

\title{An Intermittent Model for the 1/f Spectrum in the Pristine Solar Wind}% Force line breaks with \\

\author{M. Brodiano}
\affiliation{Universidad de Buenos Aires, Facultad de Ciencias Exactas y Naturales, Departamento de Física, Ciudad Universitaria, 1428 Buenos Aires, Argentina.}
\affiliation{CONICET - Universidad de Buenos Aires, Instituto de Física Interdisciplinaria y Aplicada (INFINA), Ciudad Universitaria, 1428 Buenos Aires, Argentina.}

\author{F. Sahraoui}
\affiliation{Laboratoire de Physique des Plasmas (LPP), CNRS, École Polytechnique, Sorbonne Université, Université Paris-Saclay, Observatoire de Paris, 91120 Palaiseau, France}

\author{D.Manzini}
\affiliation{Astronomy Unit, School of Physics and Astronomy, Queen Mary University of London, London E1 4NS, UK}

\author{L.Z.Hadid}
\affiliation{Laboratoire de Physique des Plasmas (LPP), CNRS, École Polytechnique, Sorbonne Université, Université Paris-Saclay, Observatoire de Paris, 91120 Palaiseau, France}

%\affiliation{Laboratoire de Physique des Plasmas (LPP), CNRS, École Polytechnique, Sorbonne Université, Université Paris-Saclay, Observatoire de Paris, 91120 Palaiseau, France}
%\affiliation{Dipartimento di Fisica E.Fermi, University of Pisa, Italy}

\author{F.Pugliese}
\affiliation{Universidad de Buenos Aires, Facultad de Ciencias Exactas y Naturales, Departamento de Física, Ciudad Universitaria, 1428 Buenos Aires, Argentina.}
\affiliation{CONICET - Universidad de Buenos Aires, Instituto de Física Interdisciplinaria y Aplicada (INFINA), Ciudad Universitaria, 1428 Buenos Aires, Argentina.}

\author{P. Dmitruk}
\affiliation{Universidad de Buenos Aires, Facultad de Ciencias Exactas y Naturales, Departamento de Física, Ciudad Universitaria, 1428 Buenos Aires, Argentina.}
\affiliation{CONICET - Universidad de Buenos Aires, Instituto de Física Interdisciplinaria y Aplicada (INFINA), Ciudad Universitaria, 1428 Buenos Aires, Argentina.}
\affiliation{Institut Franco-Argentin de Dynamique des Fluides pour l’Environnement (IFADyFE), Universidad de Buenos Aires, CONICET, CNRS (IRL 2027). Buenos Aires, Argentina}

\author{N. Andr\'es}
\affiliation{Universidad de Buenos Aires, Facultad de Ciencias Exactas y Naturales, Departamento de Física, Ciudad Universitaria, 1428 Buenos Aires, Argentina.}
\affiliation{CONICET - Universidad de Buenos Aires, Instituto de Física Interdisciplinaria y Aplicada (INFINA), Ciudad Universitaria, 1428 Buenos Aires, Argentina.}
\affiliation{Institut Franco-Argentin de Dynamique des Fluides pour l’Environnement (IFADyFE), Universidad de Buenos Aires, CONICET, CNRS (IRL 2027). Buenos Aires, Argentina}

\date{Last edited: \today}

\begin{abstract}
We present a statistical, observational study of the \(1/f\) range of solar wind turbulence, where \(f\) denotes frequency, using \emph{in situ} data from the Parker Solar Probe (PSP). We compute the energy cascade rate using the third‐order law of incompressible magnetohydrodynamic (MHD) turbulence, incorporating expansion terms to account for solar wind dynamics. Our results reveal a \(1/\tau\) dependence of the energy cascade rate, where \(\tau\) is the temporal lag, within the \(1/f\) range, in contrast to the constant cascade rate in the inertial range. To explain this behavior, we propose a new intermittent model predicting a \(1/\ell\) scaling of the cascade rate, where \(\ell\) represents the spatial lag. The analysis of the probability density function (PDF) of magnetic field increments confirms the intermittent nature of the parallel fluctuation component, whereas the perpendicular fluctuations are found to be quasi‐Gaussian. These findings provide new insights into energy transfer processes in the \(1/f\) range of solar wind turbulence, with potential applications in planetary magnetosheaths.
\end{abstract}

\maketitle

{\bf Introduction.}
The solar wind is a unique laboratory for studying an astrophysical plasma that is in a fully developed turbulent state thanks to the wide range of temporal and spatial scales involved \citep{BC2013,Verscharen2018}. One of the key methods for characterizing the multiscale nature of turbulence is through the power spectral density (PSD) of the turbulent fluctuations. In the solar wind, the PSD's low‐frequency part typically follows a $-1$ power‐law exponent; in contrast, within the inertial range (IR), it steepens to approximately $-5/3$ \citep[e.g.,][]{C2020}. The origin of the $1/f$ range (also known as the ``energy‐containing scales'' or the ``$1/f$ flicker noise'') remains a subject of heated debate \citep{M1986,Velli1989,Dm2007,Bemporad2008,Verdini2012,Matteini2018,C2018,Magyar2022,Huang2023,Chen_2020}. Several efforts have been made to explain the emergence of the $1/f$ spectrum. Examples are the superposition of signals from uncorrelated magnetic reconnection events that occur in the corona, with their correlation times following a log‐normal distribution \citep{M1986}; the evolution of Alfvén waves originating from the corona in the expanding solar wind \citep{Velli1989,Verdini2012,Perez2013}; the nonlinear evolution of the parametric instability, which leads to an inverse cascade of the Alfvén wave quanta \citep{C2018}; and the presence of a cutoff in the distribution of the fluctuations and the saturation of their mean amplitude in Alfvénic fast streams \citep{Matteini2018}.

In this Letter, we take a different approach and attempt to answer the following questions: Is the $1/f$ range populated by fully developed turbulent fluctuations? And if so, why does it not exhibit a $-5/3$ scaling as the inertial range does, in agreement with predictions of MHD turbulence theory? To this end, we fully characterize the $1/f$ range as measured by PSP \citep{Fo2016} in the inner heliosphere, using the popular third‐order law model of incompressible MHD turbulence \citep{P1998a,P1998b,Sorriso2002,Mi2009,Bo2009,W2010,Verdini_2015,So2007,Sa2008,Co2015,Si2008,Ma2008,M1999,Mc2008,Mc2005}, quantifying the level of intermittency therein and comparing the results with those from the inertial range.

%%%%%%%%%%%%%%%%%%%%%%%%%%%%

{\bf Third‐order law of incompressible MHD turbulence.} From the MHD equations \citep[e.g.,][]{B2003}, and following the usual assumptions of time stationarity and space homogeneity, scale separation between forcing and dissipation, and infinite kinetic and magnetic Reynolds numbers \citep[e.g.,][]{A2017b}, an exact relation valid in the inertial range for incompressible MHD turbulence can be derived \citep{P1998a,P1998b}:
\begin{equation}\label{pp98}
-4\,\varepsilon \;=\; \rho_{0}\,\boldsymbol{\nabla}\cdot
\bigl\langle 
  \bigl[\bigl(\delta\mathbf{u}\bigr)^{2} 
       + \bigl(\delta\mathbf{b}\bigr)^{2}\bigr]\,
      \delta\mathbf{u} 
  \;-\; 2\,\bigl(\delta\mathbf{u}\cdot\delta\mathbf{b}\bigr)\,
             \delta\mathbf{b}
\bigr\rangle
\end{equation}
where $\varepsilon$ is the total energy cascade rate per unit volume, $\rho_{0}$ is the mean mass density, $\mathbf{u}$ is the velocity field, and $\mathbf{b}$ is the magnetic field in Alfvénic units. The operator $\delta$ denotes the field increment between two positions, $\mathbf{x}$ and $\mathbf{x}' = \mathbf{x} + \boldsymbol{\ell}$, with $\ell = \lvert\boldsymbol{\ell}\rvert$ the longitudinal distance. The angular bracket $\langle\cdot\rangle$ denotes an ensemble average, here replaced by a time average under the assumption of ergodicity \citep[e.g.,][]{B2003}. If we further assume statistical isotropy, we can integrate Eq.~\eqref{pp98} over a sphere of radius $\ell$ to obtain a scalar relation valid for isotropic turbulence:
\begin{equation}\label{CascadeEq}
-\frac{4}{3}\,\varepsilon\,\ell \;=\; 
 \rho_{0}\,\bigl\langle 
   \bigl[\bigl(\delta\mathbf{u}\bigr)^{2} 
        + \bigl(\delta\mathbf{b}\bigr)^{2}\bigr]\,
       \delta u_{\ell}
   \;-\; 2\,\bigl(\delta\mathbf{u}\cdot\delta\mathbf{b}\bigr)\,
              \delta b_{\ell}
 \bigr\rangle.
\end{equation}
When Eq.~\eqref{CascadeEq} is applied to single‐spacecraft data, the Taylor hypothesis is used to convert time lags into spatial lags, i.e., $\ell = \tau U_{0}$. The longitudinal components of the fields are then defined as $
u_{\ell} = \mathbf{u} \cdot \mathbf{\hat{U}}_{0}\quad\text{and}\quad
b_{\ell} = \mathbf{b} \cdot \mathbf{\hat{U}}_{0}$, where $\mathbf{\hat{U}}_{0}$ is the unit vector in the direction of the mean plasma flow velocity. For completeness, we add expansion terms to the right‐hand side of Eq.~\eqref{CascadeEq}. These terms, as described in \citet{Verdini2024,G2013,H2013}, account for the impact of the solar wind’s radial expansion on turbulence. In contrast to the terms in this exact relation, these additional terms are second‐order structure functions (SFs), written in terms of the velocity and magnetic fields as
\begin{equation}
    F_{\mathrm{exp}} \;=\; -\frac{V_{\mathrm{sw}}}{2R} 
    \left[ 
        \bigl\langle (\delta\mathbf{u}_{\perp})^{2} \bigr\rangle 
        \;+\; 
        \bigl\langle (\delta\mathbf{b}_{R})^{2} \bigr\rangle 
    \right],
\end{equation}
where $V_{sw}$ is solar wind velocity, $R$ is heliocentric distance, $\mathbf{u}_{\perp}$ is the component of the velocity perpendicular to the radial direction, and $\mathbf{b}_{R}$ is the radial component of the Alfvén velocity. These terms quantifies the expansion driven source term in the third order law and become relevant at large scales, where they can be comparable to the turbulent cascade rate, particularly in Alfvénic streams \citep{Verdini2024}.

%%%%%%%%%%%%%%%%%%%%%%%%%%%%

{\bf PSP Data Selection.} The magnetic field data are obtained from the fluxgate magnetometer of the FIELD suite of instruments \citep{Ba2016}, while proton density and velocity data are measured by the Solar Probe Cup of the SWEAP instrument suite \citep{K2016,C2020}. Originally, the magnetic field resolution from FIELD is 4 Hz, on average. These data were interpolated to a 30‐second cadence to match the proton density and velocity data. This ensures consistency between both data sets, which is relevant for computing the energy cascade rate in the next section. We study 18 time intervals from 4 August 2021 to 3 December 2022 (between Encounter 9 and Encounter 14), each of two days’ duration. Here, we present only two representative intervals corresponding to two different heliocentric distances: 0.16 AU and 0.44 AU. The two‐day duration was chosen to access the low‐frequency part of the spectrum where $1/f$ is likely to be observed \citep[e.g.,][]{BC2013}. We estimated the correlation time $\tau_{c}$ of the magnetic field data—defined as the time over which the correlation decreases by one e‐folding factor ($1/e \approx 0.37$) \citep{Wrench_2024,Mattaeus_2005,Cuesta_2022}—and performed a convergence test of the correlation length to ensure relative statistical stationarity of the selected samples \citep{Stawarz2022}. The obtained values of $\tau_c$ and the derived frequencies $f_c$ are shown in Table \ref{tab:table1} to compare with the spectral break marking the transition from the $1/f$ range to the inertial range.

%%%%%%%%%%%%%%%%%%%%%%%%%%%%

\begin{figure}[t!]
\centering
\includegraphics[width=\columnwidth]{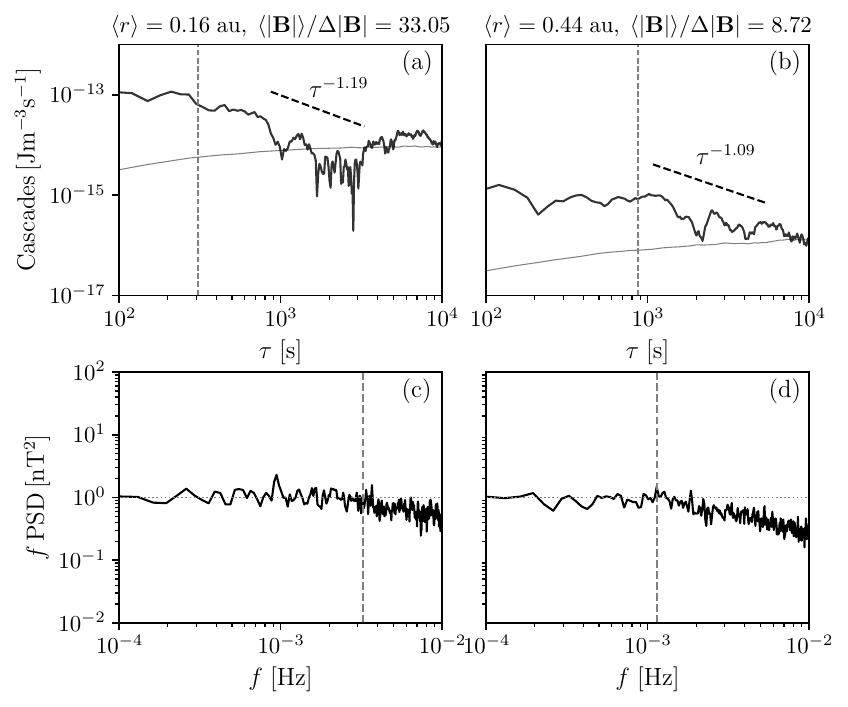}
\caption{The incompressible energy cascade rate $|\varepsilon|$ (black) and the expansion term $|\mathcal{F}_{exp}|$ (gray) as a function of time lag $\tau$ for two different heliocentric distances (a and b). In dashed black lines are the corresponding power-law fits of the total incompressible cascade rate in the $1/f$ range. Panels (c) and (d) show the corresponding compensated magnetic PSD as a function of the frequency $f$. Vertical dashed gray lines mark the correlation time $\tau_c$ (a and b) and the related correlation frequency $f_c$ (c and d). We also include information on the variation of the mean magnitude of the magnetic field with respect to its dispersion.}
\label{fig:cascade}
\end{figure}

{\bf Observational Results and an Intermittent Model.} To characterize turbulence in the $1/f$ range, we estimated the energy transfer rate as a function of time lag. Figure~\ref{fig:cascade}(a) and (b) show the full energy cascade rate, i.e., the nonlinear and expansion terms described above, and the compensated power spectral density (PSD) of the magnetic field fluctuations in panels (c) and (d). For all analyzed spectra, distinct ranges exhibiting $1/f$ scaling were observed. Within this $1/f$ range, the energy cascade rate plots exhibit a $1/\tau$ scaling (black dashed lines) before nearly flattening at $\tau < \tau_{c}$, marking the transition to the inertial range. The scale‐dependent energy transfer (or dissipation) rate reflects a \textit{non‐conservative} turbulent cascade and indicates the presence of intermittency \citep{Oboukhov_1962,Frisch_1995}.

\begin{table*}[ht]
    \centering
    \begin{tabular}{ccccccccccccc}
        \toprule[1pt]\midrule[0.3pt]
           Start time & End time &$\langle r \rangle$ [au] & $U_0$ [km/s] & $\delta u_0$ [km/s] & $U_A$ [km/s] & $M_A$ & $\tau_c$ [s] & $f_c$ [Hz]\\
        \noalign{\smallskip}
        \hline
        \noalign{\smallskip}
        29-05-2022 & 31-05-2022 & 0.16 & 425.01 & 48.93 & 293.33 & 1.45 & 308 & $3.2\times 10^{-3}$\\
        12-02-2022 & 14-02-2022 & 0.44  & 411.45 & 42.04 & 83.48 & 4.93 & 870 & $1.1\times 10^{-3}$ \\
       \midrule[0.3pt]\bottomrule[1pt]
    \end{tabular}
    \caption{From left to right: start and end time of the events, mean value of the heliocentric distance, mean plasma flow speed, rms value of the outer-scale (energy-containing range) fluid velocity, Alfv\'en speed, Alfvén Mach number and the correlation time.}
    \label{tab:table1}
\end{table*}

To provide an explanation for this new scaling law of the energy cascade rate, we turn to classical dimensional analysis. Using Eq.~\eqref{CascadeEq} and assuming $\delta\mathbf{u}\sim\delta\mathbf{b}\sim\delta u_{\ell}\sim\delta b_{\ell}$, one obtains in the inertial range
\begin{equation}
\label{eq_sc1}
    S_{3}(\ell) = \langle \delta b_{\ell}^{3}\rangle \sim \varepsilon\,\ell,
\end{equation}
where $\varepsilon$ is assumed constant. One then readily obtains the Kolmogorov spectrum $b_{k}^{2}\sim k^{-5/3}$ from relation \eqref{eq_sc1}. Now, we introduce a scale‐dependent dissipation rate $\varepsilon_{\ell}$ and assume that the Kolmogorov third‐order law \eqref{eq_sc1} remains valid in the $1/f$ range. This can be viewed as a tentative extension of the refined self‐similarity hypothesis to the $1/f$ range (see discussion below). We then write
\begin{equation}
\label{eq_sc2}
    \delta b_{\ell} \sim (\varepsilon_{\ell}\,\ell)^{1/3}.
\end{equation}
Assuming the cascade rate $\varepsilon_{\ell}$ obeys a power‐law scaling\footnote{This assumption allows for obtaining the scaling law \eqref{eq_sc3} directly by integrating Eq.~\eqref{pp98}.}, namely $\varepsilon_{\ell} \sim \ell^{\alpha}$, and introducing this relation into Eq.~\eqref{eq_sc2}, one obtains $\delta b_{\ell} \sim \ell^{(1+\alpha)/3}$. From this relation and following the usual steps of dimensional analysis, one obtains the following scaling of the magnetic energy spectrum:
\begin{equation}
\label{eq_sc3}
   b_{k}^{2} \sim k^{-2(1+\alpha)/3 - 1}.
\end{equation}
From relation \eqref{eq_sc3}, it is straightforward to infer that if the magnetic energy spectrum scales as $b_{k}^{2} \sim k^{-1}$, then $\alpha = -1$, which in turn yields a scale‐dependent energy dissipation rate:
\begin{equation}
\label{eq_sc4}
   \varepsilon_{\ell} \sim \ell^{-1}.
\end{equation}

This prediction is consistent with the estimated dissipation rates in Fig.~\ref{fig:cascade} (with $\ell = U_{0}\,\tau$, where we assume the Taylor hypothesis). To the best of our knowledge, this is the first intermittent model proposed to explain the $1/f$ range of solar wind turbulence. While several models (generally inspired by studies in hydrodynamic turbulence) were used to study intermittency at small scales (i.e., scales belonging to the IR), our model and observations address intermittency in the $1/f$ range. Note also that intermittency is tackled here through the scale‐dependent energy cascade (or dissipation) rate, while it is generally analyzed through the field increments \citep{Frisch_1995}.

The scale‐dependent cascade rate $\varepsilon_{\ell} \sim 1/\ell$ implies a non‐conservative cascade: large scales process energy at slower rates compared to small ones. In the (mathematical) limit $\ell \rightarrow +\infty$, $\varepsilon_{\ell} \rightarrow 0$. To provide a physical interpretation of this result, we assume the existence of an integral scale $L_{0}$ as an upper bound on $\ell$, to which we associate a cascade rate $\varepsilon_{L_{0}} \sim 1/L_{0}$. In the small‐scale limit, we assume that turbulence forms an IR at the correlation scale $L_{c}$, where the energy is transferred at a constant rate $\varepsilon_{\mathrm{IR}}$ (i.e., a conservative cascade). Within these assumptions, the $1/f$ range is bound to process energy at a rate that must fulfill two boundary conditions: $\varepsilon_{\ell} = \varepsilon_{L_{0}}$ for $\ell = L_{0}$ and $\varepsilon_{\ell} = \varepsilon_{\mathrm{IR}}$ for $\ell = L_{c}$. These conditions then set the mathematical form (e.g., the slope) of the cascade rate in the $1/f$ range. Thus, we speculate that the scaling of the cascade rate in the $1/f$ range may not be universal. Instead, the scaling would depend on the system size—namely, the ratio between the integral and correlation lengths, $L_{0}/L_{c}$. This scenario might explain the variability of magnetic PSD slopes observed in planetary magnetosheaths on the largest scales, where spectral indices other than $f^{-1}$ have frequently been reported \citep{Hadid_2015,Huang_2017}.

%%%%%%%%%%%%%%%%%%%%%%%%%%%%

\begin{figure}[t!]
\centering
\includegraphics[width=\columnwidth]{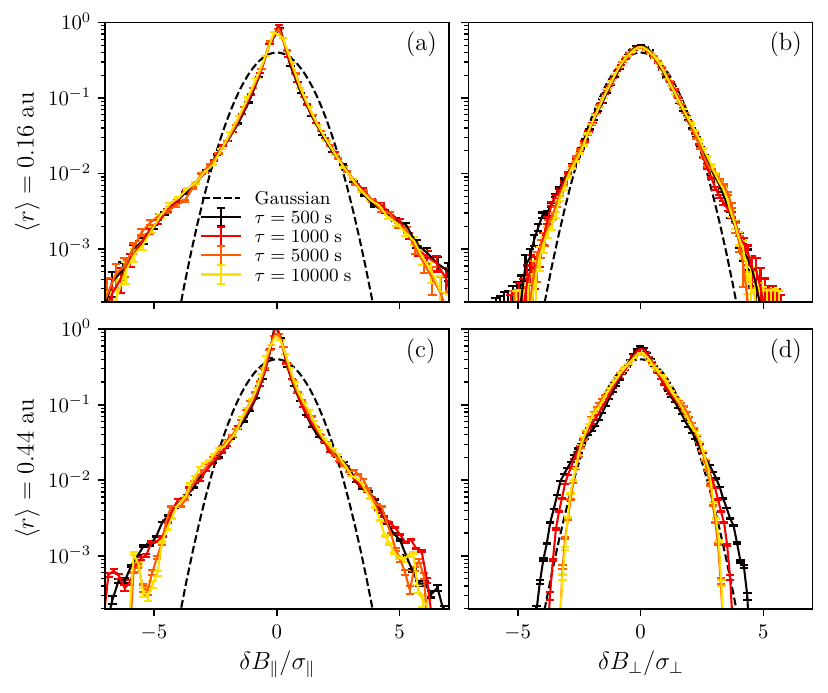}
\caption{PDFs of the parallel and perpendicular magnetic field increments for $\tau = \{500,\,1000,\,5000,\,10000\}\,\mathrm{s}$ in the $1/f$ range, at heliocentric distances of 0.16\,AU (panels a and b) and 0.44\,AU (panels c and d). A Gaussian distribution (black dashed curve) is shown for comparison.}
\label{fig:PDF}
\end{figure}

{\bf Observational Results and Intermittency in the $1/f$ Range.} To characterize intermittency in the $1/f$ range, we estimate the probability density function (PDF) of the magnetic field fluctuations, which we decompose into parallel and perpendicular components using a sliding‐window approach \citep[e.g.,][]{Huang_2014}. The mean magnetic field within each window is defined as the parallel direction, while the two perpendicular components are determined through projection. The window width is selected based on the stationarity condition, ensuring compliance within a specified tolerance. Figure~\ref{fig:PDF} shows the PDFs (normalized to their standard deviation) of the parallel and perpendicular components of the magnetic field increments in the $1/f$ range for the same two samples taken at different heliocentric distances. For comparison, a Gaussian distribution (dashed lines) is overplotted. In all cases, the PDFs of the parallel component deviate significantly from a Gaussian distribution, indicating high levels of intermittency \citep{M2015,Hadid_2015}. In contrast, the PDFs of the perpendicular increments exhibit a quasi‐Gaussian distribution, indicating their random‐like nature. This suggests that the parallel fluctuations might have had “enough” time to evolve nonlinearly, as if they emerged deeper in the corona, compared to the perpendicular ones. The present observations are at odds with the known properties of turbulence in the IR of the solar wind, where both parallel and perpendicular fluctuations exhibit highly non‐Gaussian statistics \citep{Kiyani2013}. They highlight the leading role that appears to be played by the (non‐symmetric) heavy tails of the PDF of parallel fluctuations in driving the scale‐dependent transfers in the $1/f$ range. As the cascade proceeds to small scales close to the IR, the PDFs of the perpendicular fluctuations become heavy‐tailed (not shown) and contribute more effectively to the cascade, hence the increasing transfer rate until reaching its upper bound $\varepsilon_{\mathrm{IR}}$ at the scale $L_{c}$. 

In our data set, as in the pristine solar wind in general, the parallel fluctuations are sub‐dominant (they represent only $\sim 20\mbox{--}30\%$ of the total fluctuations \citep{Howes_2014}). Nevertheless, they seem to drive most of the transfers in the $1/f$ range. This apparent contradiction might be removed by observing that the longitudinal increments $\delta u_{\ell}$ and $\delta u_{A\ell}$ in Eq.~\eqref{CascadeEq} are quasi‐parallel/anti‐parallel (or moderately oblique) to $\mathbf{B}$, considering average angles $\theta_{vB}=35^\circ$ and $45^\circ$ for cases at 0.16 au and 0.44 au, respectively. Thus, the 1D sampling of the turbulence using the Taylor hypothesis allows capturing the role of the parallel fluctuations. Another remark is that the incompressible model we used seems to capture the role of the parallel (compressible) fluctuations. This may indicate that the compressible fluctuations are “slaved” to the incompressible ones \citep{Schekochihin_2009}. We performed an estimation of the compressible cascade rate, using the isothermal model of \citet{Andres_2018}, and found no significant changes to the cascade rates in Fig.~\ref{fig:cascade} (minor changes were found at the largest scales of the case at 0.16 au).

\begin{figure}[t!]
\centering
\includegraphics[width=0.99\columnwidth]{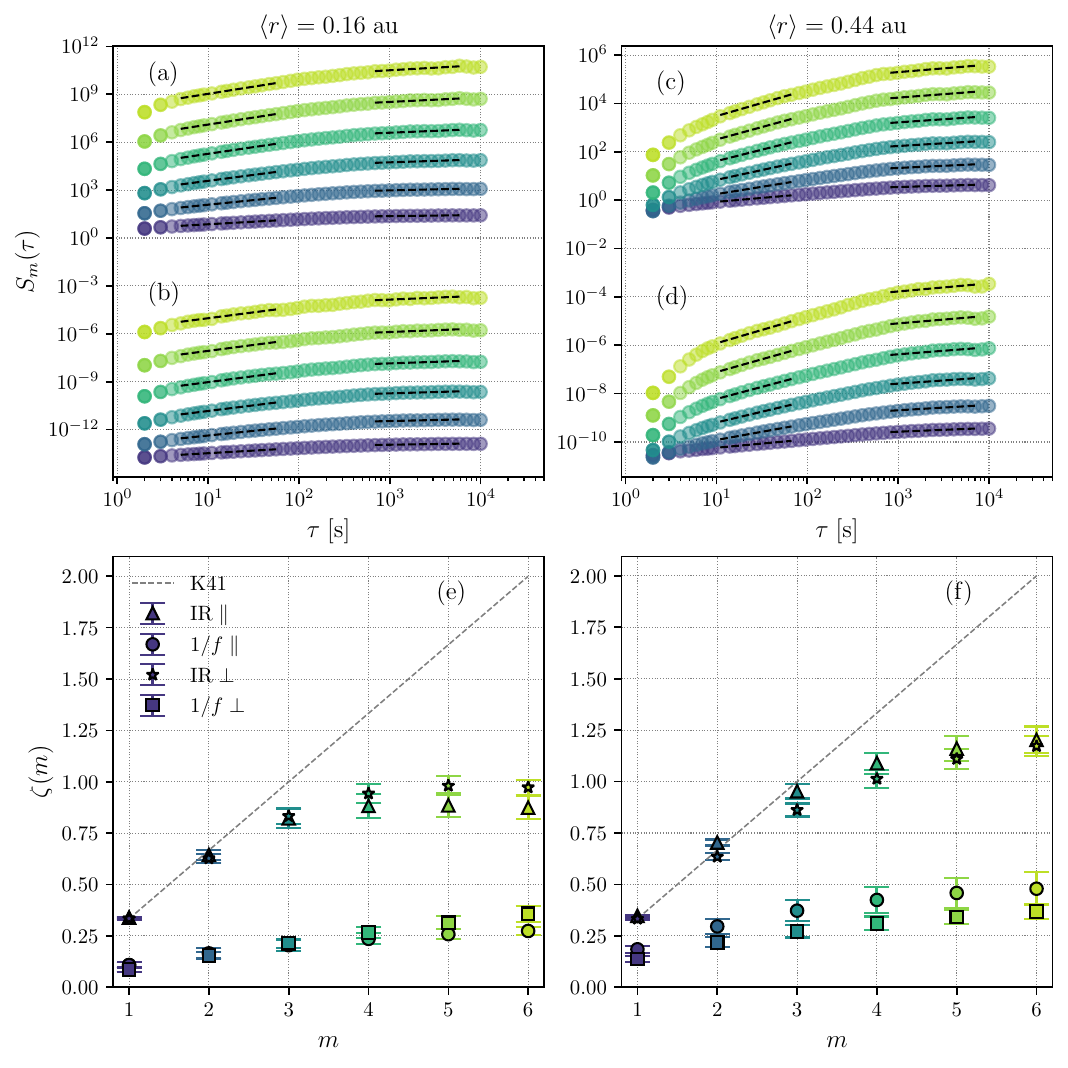}
\caption{Top: Structure functions of the perpendicular (a and c) and parallel (b and d) fluctuations, $S_{m}(\tau) = \langle \lvert B_{\perp,\parallel}(t+\tau) - B_{\perp,\parallel}(t)\rvert^{m}\rangle$, as a function of the time lag $\tau$ for different orders $m$, spanning both the $1/f$ and inertial ranges. Bottom: the scaling exponent $\zeta(m)$ as a function of the order, obtained from $S_{m}(\tau)\propto\tau^{\zeta(m)}$, in the inertial range (triangle and star markers) and the $1/f$ range (circle and square markers). The gray‐dashed line indicates the prediction from Kolmogorov theory (K41), $\zeta(m)=m/3$, for self‐similar turbulent flows.}
\label{fig:SF}
\end{figure}

\begin{figure}[t!]
\centering
\includegraphics[width=\columnwidth]{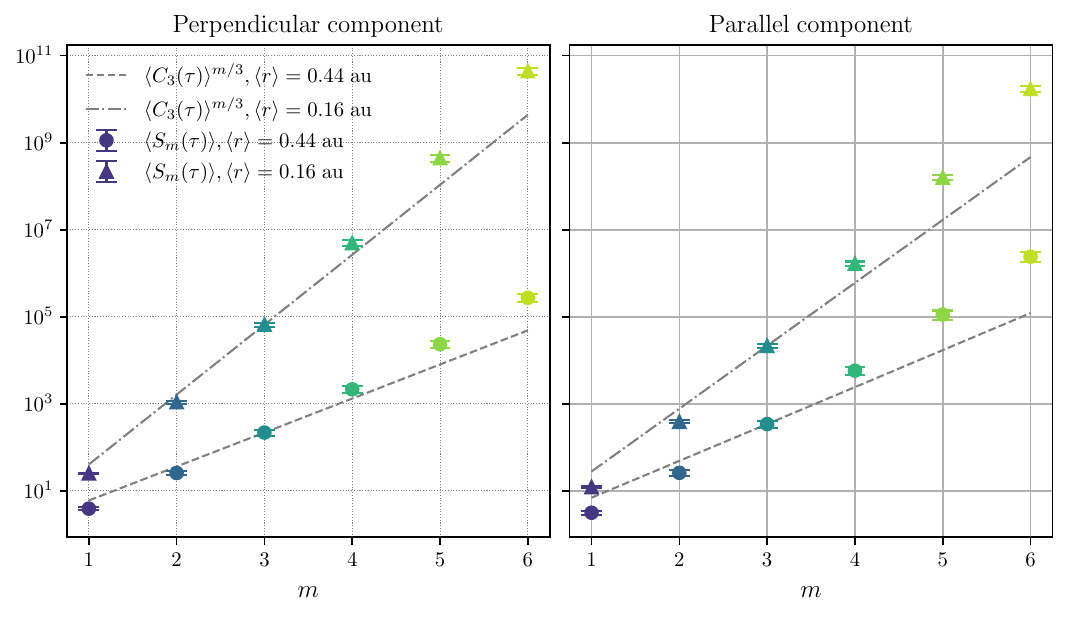}
\caption{Estimation of the average constant $C_{m}(\tau)$ in the $1/f$ range (error bars denote one standard deviation) for different orders $m$ of the structure functions (dots), compared to the prediction $C_{m} = C_{3}^{\,m/3}$ (gray dashed line) if self‐similarity is achieved in the $1/f$ range.}
\label{fig:Sm}
\end{figure}

In the IR, the self‐similarity hypothesis of the Kolmogorov theory \citep{K41} allows us to generalize Eq.~\eqref{eq_sc1} to any order $m$, namely 
\[
S_{m}(\ell)\sim (\varepsilon\,\ell)^{m/3}.
\]
Assuming this hypothesis to be valid for the scale‐dependent transfer rate $\varepsilon_{\ell}$ in the $1/f$ range, we can generalize $S_{3}(\ell)=C_{3}\,\varepsilon_{\ell}\,\ell$ to any order $m$, namely
\begin{equation}
\label{eq_sc5}
   S_{m}(\ell)=(C_{3}\,\varepsilon_{\ell}\,\ell)^{m/3} \;=\; C_{m}\,(\varepsilon_{\ell}\,\ell)^{m/3},
\end{equation}
where we introduced the constants (i.e., scale‐independent) $C_{m} = C_{3}^{\,m/3}$ for the purpose of this study. Since the prediction $\varepsilon_{\ell}\sim 1/\ell$ was obtained from $S_{3}$, which thus acts as a stringent ``boundary condition,'' we infer that $S_{3}(\ell)$ should form a plateau in the $1/f$ range at a value $C_{3}$ whose numerical value can be fixed from the data. Likewise, Eq.~\eqref{eq_sc5}, based on the (refined) self‐similarity assumption, i.e., $\varepsilon_{\ell}^{\,m/3}\sim \ell^{-\,m/3}$, implies that higher‐order structure functions (SFs) should all form a plateau in the $1/f$ range at values $C_{m} = C_{3}^{\,m/3}$. Alternatively, any departure from that prediction for higher‐order SFs would mean that the dissipation rate, taken as a random variable, is not scale‐invariant. This prediction is readily tested in Fig.~\ref{fig:SF}, plotted for both cases at 0.16 AU and 0.44 AU for the perpendicular (a and c) and parallel (b and d) fluctuations. We observe indeed (at the top of Fig.~\ref{fig:SF}) a weak scale‐dependence of SFs in the $1/f$ range, in contrast with the scaling in the inertial range. Assuming power‐law dependence of the SFs, 
\[
S_{m}(\tau)\sim \tau^{\zeta(m)},
\]
one can extract the scaling exponents $\zeta(m)$, which are plotted at the bottom of Fig.~\ref{fig:SF}. First, in the IR, we recover known results that turbulence (both for parallel and perpendicular fluctuations) is multifractal, as manifested by the clear departure from the Kolmogorov prediction for self‐similar flows, $\zeta(m)=m/3$ for $m>3$. In the $1/f$ range, we find a weaker dependence of the scaling exponents on $m$ for both parallel and perpendicular fluctuations: the dependence is the weakest for the latter, which likely originates from their quasi‐Gaussian statistics seen in Fig.~\ref{fig:PDF}. Nevertheless, the dependence of $\zeta(m)$ on $m$ in the $1/f$ range (when statistical error bars are considered) seems to be physical and reflects a scale‐dependence of the high‐order SFs and the violation of self‐similarity. This is further confirmed by testing the prediction $C_{m} = C_{3}^{\,m/3}$. The results are shown in Fig.~\ref{fig:Sm}. We observe a slight departure from the predicted values of the constant $C_{3}^{\,m/3}$ for increasing order $m$, indicating a violation of the self‐similarity hypothesis used to derive Eq.~\eqref{eq_sc5}.

The intermittent nature of the $1/f$ range is revealed here through the scale‐dependent (mean) transfer rate and the constants $C_{m}$. An alternative approach would be to consider the dissipation rate $\varepsilon_{i}(\ell)$ as a random variable that reflects different realizations of the flow. Here, we deal only with its mathematical expectation: $\varepsilon_{\ell} \;=\; \langle \varepsilon_{i}(\ell)\rangle$. Its intermittent character can be further quantified via its moments, $\varepsilon_{\ell}^{(m)} \;=\; \langle \varepsilon_{i}^{m}(\ell)\rangle$.

%%%%%%%%%%%%%%%%%%%%%%%%%%%%

{\bf Conclusions.} In this Letter, we present new observations of the $1/f$ range of solar wind turbulence from PSP data. The first result is that the $1/f$ scaling of the magnetic energy spectra in that range reflects a non‐conservative cascade at a rate $\varepsilon_{\ell} \sim 1/\ell$ (assuming the validity of the Taylor hypothesis at those scales). The key assumption that allowed us to draw this conclusion is the extension of the third‐order law’s validity to the $1/f$ range. The main caveat is that the notion of scale separation (i.e., locality) used to derive Eq.~\eqref{CascadeEq} might not hold on the largest scales involved in the $1/f$ range. The second key result is that energy transfer in the $1/f$ range is intermittent and is likely driven by parallel fluctuations. Some evidence of departure from (refined) self‐similarity in the $1/f$ range, based on structure functions of the magnetic field, is provided. However, a more rigorous intermittent model based on moments of the dissipation rate remains to be developed. Overall, these results call for revisiting the physics of the $1/f$ range in the solar wind and planetary magnetosheaths, and for extending intermittency models of turbulence to scales much larger than those of the inertial range.

\begin{acknowledgments}
All authors acknowledge financial support from the ECOS SUD 2022 $\#$ A22U02 CNRS / CONICET grant. M.B., F.P., P.D.~and N.A.~acknowledges financial support from the following grants: PIP Grant No.~11220200101752, UBACyT Grant No.~20020220300122BA and Redes de Alto Impacto REMATE from Argentina. L.H.Z., D.M. and N.A.~acknowledges support by the International Space Science Institute (ISSI) in Bern, through ISSI International Team project $\#$ 23-591. D.M.~is supported by STFC Consolidated Grant ST/X000974/1. The data supporting the findings of this study are openly available in NASA Parker Solar Probe Data Repository \citep{1}. \end{acknowledgments}

%\appendix

\bibliographystyle{apsrev4-2}
\bibliography{main}

\end{document}